\documentclass[fleqn,usenatbib]{mnras}

\usepackage{newtxtext,newtxmath}
\usepackage[T1]{fontenc}
\usepackage[version=4]{mhchem}

\usepackage{dcolumn}% Align table columns on decimal point
\usepackage{bm}% bold math
\usepackage{etoolbox}
\usepackage{threeparttable}
\usepackage{array}
\usepackage{xspace}
\usepackage{multirow}
\usepackage{booktabs}
\usepackage[version=4]{mhchem}
\usepackage{xcolor}
\usepackage{natbib}

\newcolumntype{d}{D{.}{.}{5}}
\newcolumntype{e}{D{.}{.}{10}}
\newcommand{\wn}{cm$^{-1}$\xspace} % wavenumbers
\newcommand{\mrm}[1]{\ensuremath{\mathrm{#1}}}

\DeclareRobustCommand{\VAN}[3]{#2}
\let\VANthebibliography\thebibliography
\def\thebibliography{\DeclareRobustCommand{\VAN}[3]{##3}\VANthebibliography}

\usepackage{graphicx}	% Including figure files
\usepackage{amsmath}	% Advanced maths commands
\newcommand{\bsy}[1]{\ensuremath{\boldsymbol{#1}}}

\title[\ce{HC^{17}O+}--\ce{H2} collisions]{Hyperfine resolved rate coefficients of \ce{HC^{17}O+} with \ce{H2} $\bsy{(j = 0)}$} 

\author[F. Tonolo et al.]{
F. Tonolo,$^{1,2}$
F. Lique,$^{3}$
M. Melosso,$^{4}$ 
C. Puzzarini,$^{2}$
and L. Bizzocchi$^{2}$
\thanks{E-mail: luca.bizzocchi@unibo.it}
\\
$^{1}$Scuola Normale Superiore, Piazza dei Cavalieri 7, I-56126 Pisa, Italy.\\
$^{2}$Dipartimento di Chimica ``Giacomo Ciamician'', Universit\`a di Bologna, Via F. Selmi 2, I-40126 Bologna, Italy.\\
$^{3}$Univ. Rennes, CNRS, IPR (Institut de Physique de Rennes) -- UMR 6251, F-35000 Rennes, France.\\
$^{4}$Scuola Superiore Meridionale, Largo San Marcellino 10, 80138 Naples, Italy.
}
\date{Accepted XXX. Received YYY; in original form ZZZ}

\pubyear{2022}

\begin{document}
\label{firstpage}
\pagerange{\pageref{firstpage}--\pageref{lastpage}}
\maketitle

\begin{abstract}
The formyl cation (\ce{HCO+}) is one of the most abundant ions in molecular clouds and  plays a major role in the interstellar chemistry. 
For this reason, accurate collisional rate coefficients for the rotational excitation of \ce{HCO+} and its isotopes due to the most abundant perturbing species in interstellar environments are crucial for non-local thermal equilibrium models and deserve special attention.
In this work, we determined the first hyperfine resolved rate coefficients of \ce{HC^{17}O+} in collision with \ce{H2} $(j=0)$. Indeed, despite no scattering calculations on its collisional parameters have been performed so far, the \ce{HC^{17}O+} isotope assumes a prominent role for astrophysical modelling applications. 
Computations are based on a new four dimensional (4D) potential energy surface, obtained at the CCSD(T)-F12a/aug-cc-pVQZ level of theory. A test on the corresponding cross section values pointed out that, to a good approximation, the influence of the coupling between rotational levels of $\ce{H2}$ can be ignored. For this reason, the \ce{H2} collider has been treated as a spherical body and an average of the potential based on five orientations of \ce{H2} has been employed for scattering calculations. 
State-to-state rate coefficients resolved for the \ce{HC^{17}O+} hyperfine structure for temperature ranging from 5 to 100\,K have been computed using recoupling techniques.
This study provides the first determination of \ce{HC^{17}O+}--\ce{H2} inelastic rate coefficients directly computed from full quantum close-coupling equations, thus supporting the reliability of future radiative transfer modellings of \ce{HC^{17}O+} in interstellar environments. 

\end{abstract}


%%%%%%%%%%%%%%%%%%%%%%%%%%%%%%%%%%%%%%%%%%%%%%%%%%
%%%%%%%%%%%%%%%%% BODY OF PAPER %%%%%%%%%%%%%%%%%%
%%%%%%%%%%%%%%%%%%%%%%%%%%%%%%%%%%%%%%%%%%%%%%%%%%

\section*{Introduction}
\indent\indent
When aiming at a deeper knowledge of the physical and chemical properties of star forming regions, the formyl cation (\ce{HCO+}) and its isotopic variants undoubtedly represent one of the most attracting systems for a number of reasons.
\ce{HCO+} is widespread and abundant through many evolutionary stages of the interstellar medium (ISM) and its formation path is straightforward and well understood \citep{herbst1973formation}. 
Due to its ionic nature, it plays a major role both in the interstellar chemistry and the dynamics of the interstellar gas. It also provides much information on the characteristics of the sources responsible for its heating and ionization \citep{guelin1982state,jorgensen2004molecular}.
Moreover, its main destruction route occurs by dissociative recombination, making its abundance closely related to the electron density of the studied region \citep{wootten1979determination}. 

Generally speaking, the capability of modelling the collisional excitation is critical to derive reliable molecular abundance from the emission spectra of interstellar clouds \citep{roueff2013molecular}.
This is particularly true for species that are abundant in the ISM since their emission lines are often optically thick.
To this aim, many efforts have been made in order to retrieve accurate state-to-state collisional rate coefficients for \ce{HCO+} interacting with the \ce{He} and \ce{H2} \citep{masso2014hco+,yazidi2014revised,tonolo2021improved}, as well as for some of its isotopologues (see e.g., \citealt{pagani2012method}, \citealt{denis2020rotational} and references therein).

However, in spite of several spectroscopic studies and astrophysical detections \citep[see e.g.,][]{guelin1982detection,plummer1983laboratory,dore2001laboratory}, to the best of our knowledge, a thorough study of the collision physics of the \ce{HC^{17}O+}$-$\,\ce{H2} system has not been carried out yet.
The rotational transitions of the \ce{^{17}O}-bearing variant exhibit hyperfine structure due to the splitting of its rotational energy levels by the electric quadrupole coupling of the \ce{^{17}O}\,$(I = 5/2)$ nucleus. 
When properly treated, the hyperfine structure is a precious source of information, as the observed intensities for the various hyperfine components provide a mean to constrain the optical depth.
Given the above motivations, the availability of up-to-date collisional data of \ce{HC^{17}O+} with the major astrophysical collision partners gains a substantial importance.

With the aim of filling this lack, we present here the first calculation of the hyperfine state-to-state rate coefficients 
for \ce{HC^{17}O+} with \ce{H2} $(j=0)$ system, based on a new four dimensional (4D) interaction potential. 
The paper is structured as follows.
In \S~\ref{comp}, the adopted computational procedure is described. In more detail, the approach employed to accurately describe the interaction potential between \ce{HC^{17}O+} and \ce{H2} is provided in \S~\ref{pes}. In \S~\ref{scat}, the computation of the collisional cross sections are described (\S~\ref{cs}), and the collisional rate coefficients are reported and analysed (\S~\ref{hf}). 
Moreover, \S~\ref{ie} assesses the impact given by isotopic substitution on the values of the computed rate coefficients.
Finally, a comparison of the rate coefficients obtained with the almost exact recoupling approach and those derived from infinite order sudden (IOS) based approximation \citep{faure2012impact} and statistical methods is provided in \S~\ref{ap}, with the aim of estimating the impact of such approximations on radiative transfer applications.

\section{Computational details}\label{comp}
\subsection{Potential energy surface}\label{pes}
\indent\indent
The first step in computing the scattering parameters of a collisional system is the determination of the corresponding potential energy surface (PES). Within the Born Oppenheimer (BO) approximation, the interaction energy of the fragments does not depend on the chosen isotopologue for either the target or the collider species, the only difference is the shift of the Jacobi coordinates, used to describe the scattering, due to the variation of the centre of mass. 
To investigate the \ce{HC^{17}O+} and \ce{H2} collisional system, we thus decided to study the PES involving the parent species (\ce{HC^{16}O+}). Such a choice, allowed for a direct test of the accuracy of our methodology by comparison with previous calculations (\cite{masso2014hco+,yazidi2014revised}). 
The effect of the isotopic substitution is then  accounted for by shifting the centre of mass of the target in the PES before performing scattering calculations.

\begin{figure}
\centering
 \includegraphics[scale=0.18]{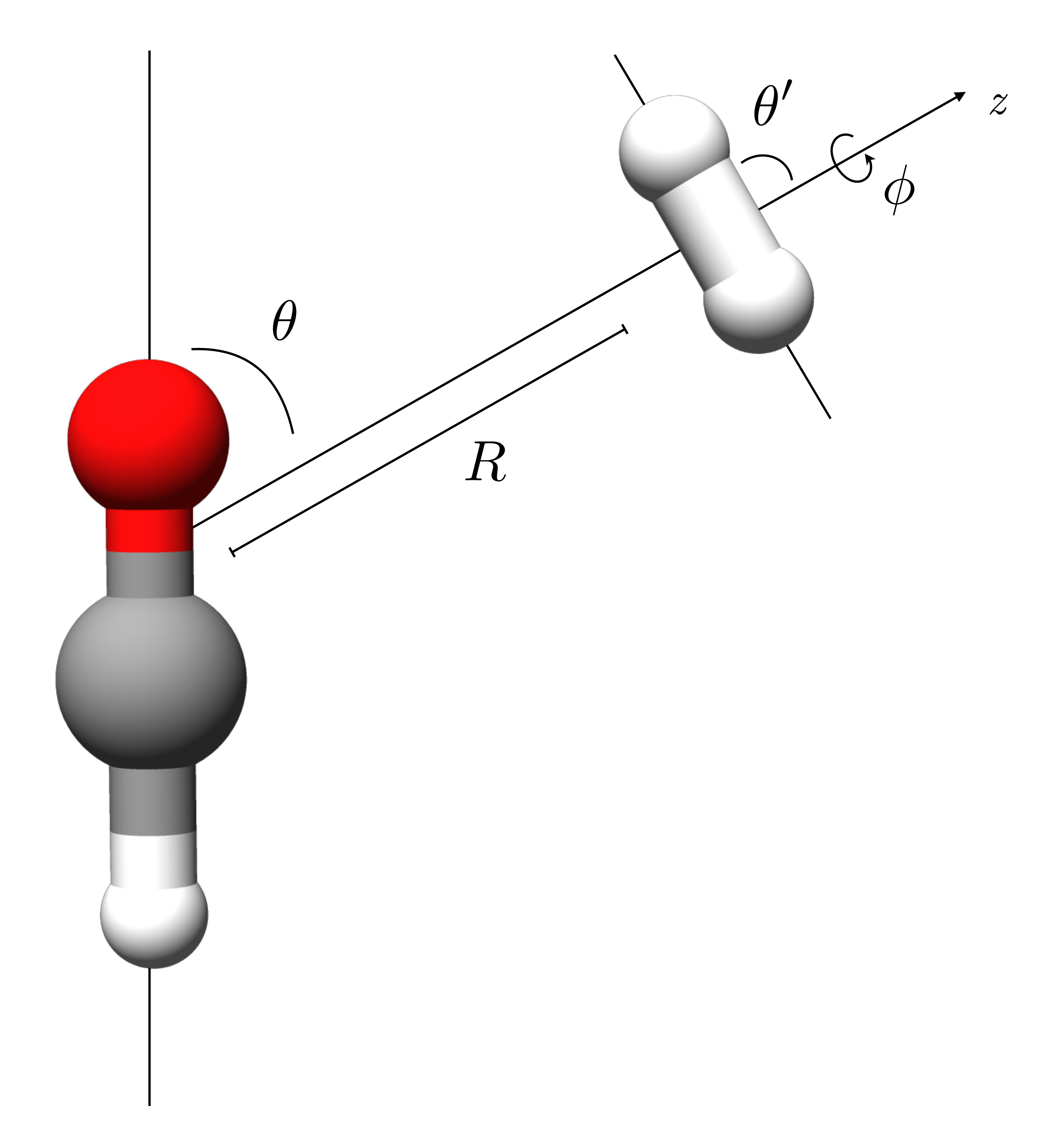}
 \caption{Jacobi internal coordinates of the \ce{HCO+} -- \ce{H2} collisional system.}
 \label{fig:geom}
\end{figure}

To describe the collisional system, a set of four Jacobi coordinates has been used (see Figure \ref{fig:geom} for a graphical representation), i.e., the $R$ distance between the centre of mass of \ce{HCO+} and that of \ce{H2}, the $\theta$ angle between the \ce{HCO+} molecular axis and the $R$ vector, and, finally, the $\theta^\prime$ and $\phi$ angles that orient the \ce{H2} molecule in the \ce{HCO+}--$R$ plane and out of it, respectively. 
\ce{HCO+} and \ce{H2} were considered as rigid rotors since, under non-reactive low-temperature conditions, all the vibrational channels are closed for all of the rotational channels taken into account. Furthermore, previous studies have shown that the rigid rotor approximation is accurate enough for low-temperature scattering applications, whenever collisional energies are lower than the bending frequency \citep{stoecklin2013ro}.
The geometry of \ce{HCO+} was held fixed at its experimentally determined equilibrium structure (linear): $r_{\textrm{CH}}= 1.0920 $\,\AA, and $r_{\textrm{CO}}= 1.1056 $\,\AA\, \citep{dore2003study}. 
For \ce{H2}, we used the bond length $r_{\textrm{HH}}= 0.7667 $\,\AA, which corresponds to the vibrationally averaged value of its ground state \citep{jankowski1998ab}.

The interaction energies between \ce{HCO+} and \ce{H2} have been computed by employing an explicitly correlated method, denoted by the -F12 suffix \citep{adler2007simple,knizia2009simplified,peterson2008systematically}, based on the coupled cluster theory and accounting for the singles, doubles and a perturbative treatment of triples excitations \citep[CCSD(T),][]{raghavachari1989fifth}. 
This has been combined with a correlation consistent quadruple-$\zeta$ quality basis set, thus leading to the CCSD(T)-F12/aug-cc-pVQZ level of theory \citep{dunning1989gaussian}. 
This choice was supported by the demonstrated accuracy of explicitly correlated methods in  describing long- and short-range multi-dimensional PESs \citep{lique2010benchmarks,ajili2013accuracy}. Moreover, the addition of diffuse functions in the basis set improves the description of the electronic behaviour for molecular systems involving non-covalent interactions \citep{kendall1992electron,lupi2021junchs}. 
All calculations were carried out with the MOLPRO suite of programs \citep{werner2012wires}.

The interaction energy $E_{\textrm{int}}$ has been computed as:
\begin{equation}
 E_{\text{int}} = E_\mrm{AB} - (E_\mrm{A} + E_\mrm{B}) \,,
\end{equation}
where $E_\mrm{AB}$ is the molecular complex energy, while $E_\mrm{A}$ and $E_\mrm{B}$ are the energies of the two fragments (here generally denoted as $A$ and $B$). 
Each energy has also been corrected for the basis set superposition error (BSSE) by means of the counterpoise (CP) correction scheme.
The CP correction is computed using the \citet{boys1970calculation} scheme:
\begin{equation}
 \Delta E_{\text{CP}}= (E^{\mrm{AB}}_{\mrm{A}} - E^{\mrm{A}}_{\mrm{A}}) + (E^{\mrm{AB}}_{\mrm{B}} - E^{\mrm{B}}_{\mrm{B}})\,,
\end{equation}
where $E^{\mrm{AB}}_{\mrm{X}}$ is the energy of the monomer calculated with the same basis set used for the cluster and $E^{\mrm{X}}_{\mrm{X}}$ is the energy of the monomer computed with its own basis set ($\mrm{X} = \mrm{A},\mrm{B}$). 

\begin{figure*}
 \includegraphics[scale=0.63]{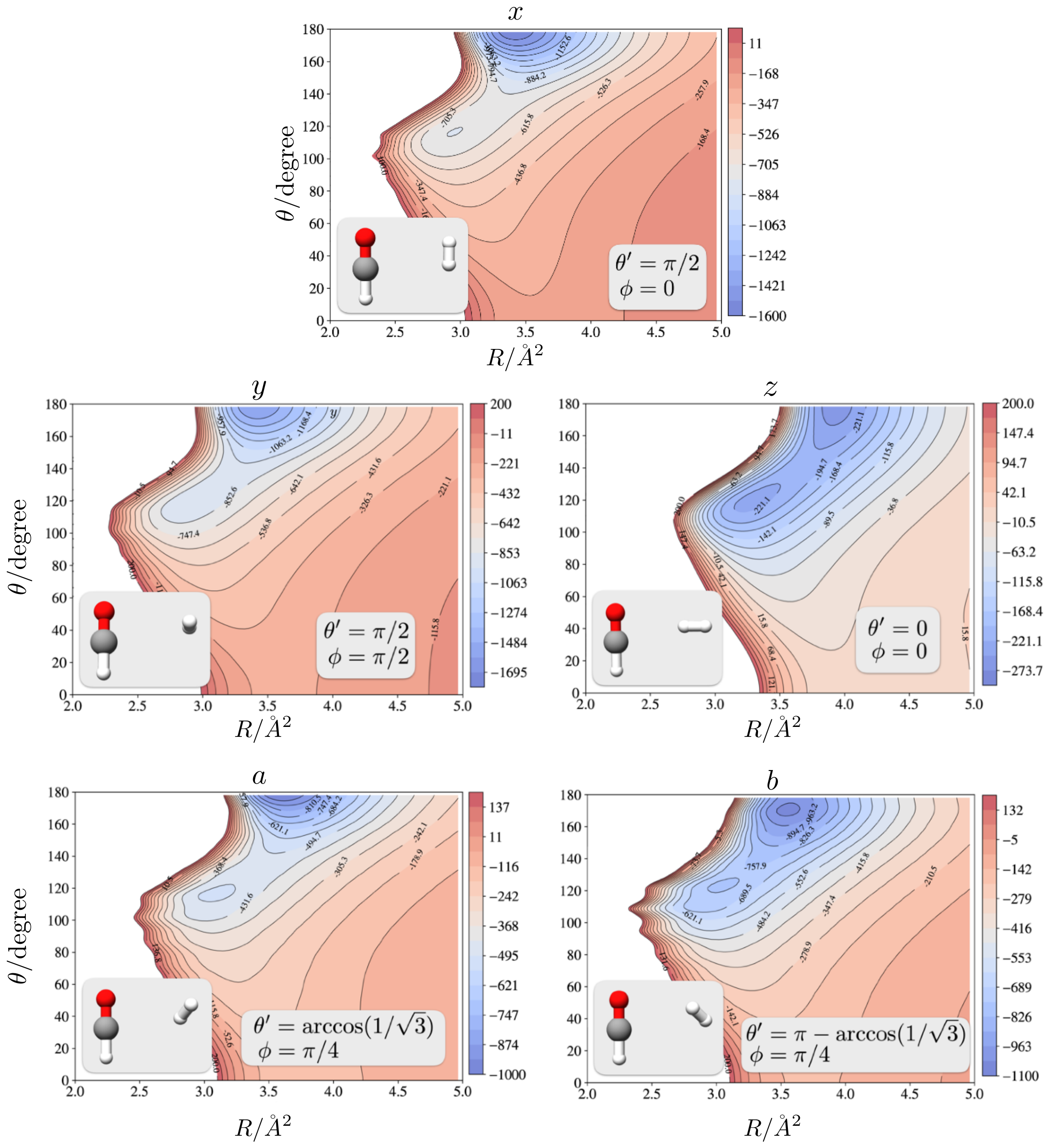}
 \caption{Contour plots of the \ce{HCO+} -- \ce{H2} interaction PES for five different orientations of \ce{H2} with respect to \ce{HCO+} as a function of $R$ and $\theta$. Energies are in \wn.} 
 \label{fig:pesorient}
\end{figure*}

We mapped the PES of the collisional system by selecting an irregular grid of 3375 points in the $\{R,\theta,\theta^\prime,\phi\}$ set of coordinates. We chose 25 values of the $\theta$ angle, equally spaced from 0 to 180 degrees, and 27~$R$ distances, which vary between 2.5 and 10\,\AA, with a denser mesh in the proximity of the potential well. Finally, for each set of $R$ and $\theta$, five different orientations of \ce{H2} with respect to \ce{HCO+} ($\theta'$ and $\phi$ angles) have been selected. The use of such a limited subset of target-collider orientations is discussed more in detail in the following paragraphs.

To solve the close-coupling equations, a suitable angular expansion of the PES for each inter-molecular distance $R$ is needed. For the \ce{HCO+}$-$\,\ce{H2} collisional system, this expansion assumes the form of an interaction potential between two linear rigid rotors \citep{green1975rotational,wernli2007rotational,wernli2007rotationalb}:
\begin{equation}\label{fit}
 V\left(R,\theta,\theta',\phi\right) = \sum_{l_1 l_2 \mu} v_{l_1 l_2 \mu}(R) s_{l_1 l_2 \mu}\left(\theta,\theta',\phi\right) \,,
\end{equation}
where $v_{l_1 l_2 \mu}(R)$ are the radial coefficients and the $l_1$, $l_2$ and $\mu$ indices are associated with the rotational angular momenta of \ce{HCO+}$(j_1)$, \ce{H2}$(j_2)$, and with their vector sum, respectively. In Eq.~\eqref{fit}, the centrosymmetric nature of \ce{H2} forces the index $l_2$ to be even. 
The $s_{l_1 l_2 \mu}$ coefficients are defined as products of spherical harmonics ($Y_{l_i m}(\theta,\phi)$) and the Clebsch-Gordan vector-coupling coefficients (see \citealt{green1975rotational,edmonds2016angular,brown2003rotational} for further details):
\begin{equation}
 \begin{split}
  s_{l_1 l_2 \mu}(\theta,\theta',\phi) = \left(\frac{2l_1 + 1}{4\pi}\right)^{1/2}
    \bigg\{\langle l_1 0 l_2 0 | l_1 l_2 \mu 0\rangle P_{l_1 0}(\theta) P_{l_2 0}(\theta') \\
    +\sum_m(-1)^{m} 2\langle l_1 m l_2-m | l_1 l_2 \mu 0\rangle
    P_{l_1 m}(\theta) P_{l_2m}(\theta')\cos(m\phi)\bigg\} \,,
 \end{split}
\end{equation}
where $P_{l_i m}(\theta) = Y_{l_i m}(\theta,\phi)\mathrm{e}^{-\mathrm{i}\,m\phi}$\@.

Previous studies \citep{wernli2006improved,faure2005role} have shown that the basis terms with $l_2 > 2$ in the expansion of the potential can be safely neglected in the description of the \ce{HCO+}--\ce{H2} interaction. 
This significantly reduces the computational effort, since the only basis functions that describe the dependence of the potential on the relative orientations of \ce{H2} are the $Y_{00}$, $Y_{20}$, $Y_{21}$ and $Y_{22}$ spherical harmonics. At fixed $R$ and $\theta$ values, the dependence of the potential on the orientation of \ce{H2} is thus entirely described by the knowledge of only four orientations $\{\theta',\phi\}$ of \ce{H2}.
Hence, the choice of five sets of $\{\theta',\phi\}$ angles is not only sufficient to describe the dependence of the potential on them, but also allows for an over-determined system to test the accuracy of the $l_2$ truncation.
The five orientations chosen to describe the relative rotations of \ce{H2} with respect to the target molecule are:  
 \begin{subequations}
  \begin{align} 
  x & \rightarrow\left(\theta^{\prime}=\frac{\pi}{2}, \phi=0\right)\,; \\
  y & \rightarrow\left(\theta^{\prime}=\frac{\pi}{2}, \phi=\frac{\pi}{2}\right)\,; \\
  z & \rightarrow(\theta^{\prime}=0, \phi=0)\,; \\
  a & \rightarrow\left(\theta^{\prime}=\arccos \left(\frac{1}{\sqrt{3}}\right), \phi=\frac{\pi}{4}\right)\,; \\
  b & \rightarrow\left(\theta^{\prime}=\pi-\arccos \left(\frac{1}{\sqrt{3}}\right), \phi=\frac{\pi}{4}\right)\,.
  \end{align}
 \label{xyzab} 
 \end{subequations}
They are the same used by \citet{wernli2007rotational} for the \ce{HC3N}--\ce{H2} system.

We performed a two dimensional (2D) fit of the potential for each of the five sets of $\{\theta',\phi\}$ orientations, following the same procedure described by \citet{tonolo2021improved} for the interacting \ce{HCO+}--\ce{He} system. 
The \textit{ab initio} values and those obtained from the fitted radial coefficients of Eq.~\eqref{fit} differ by less than~$1.5\%$ across the entire grid.
Figure~\ref{fig:pesorient} illustrates the topology of the potential obtained for each orientation of \ce{H2}: only minor anisotropy of the potential with respect to the orientation of \ce{H2} is apparent, thus validating the choice to exclude the basis terms with $l_2>2$ in the angular expansion of the PES.
Each 2D cut of the potential also exhibits a minimum for $\theta\sim 180$\,degrees, wherein the hydrogen molecule approaches the hydrogen side of \ce{HCO+}. In accordance with the findings of the benchmark study of \citet{masso2014hco+} on different levels of theory, the potential well has its maximum depth when the \ce{H2} fragment is perpendicularly oriented with respect to the \ce{HCO+} plane $(\theta=\pi, \phi=\pi/2)$ and $R \sim 3.44$\,\AA\@.

Having assessed the correct behaviour of the potential for each orientation of \ce{H2}, the analytical dependence over $\{\theta',\phi\}$ was then introduced via the following equation \citep{wernli2006improved}:
\begin{equation}
\begin{aligned}
V(R, \theta_{1}, \theta^{\prime}, &\phi^{\prime})= V_{\text{av}}(R, \theta)+\\&+\frac{1}{2}\left[V(R,\theta,z)-V_{\text{av}}(R, \theta)\right]\left(3 \cos ^{2} \theta^{\prime}-1\right)+\\&+\frac{3}{2}\left[V(R,\theta,a)-V(R,\theta,b)\right] \sin \theta^{\prime} \cos \theta^{\prime} \cos \phi^{\prime}+\\&+\frac{1}{2}\left[V(R,\theta,x)-V(R,\theta,y)\right] \sin ^{2} \theta^{\prime} \cos \left(2 \phi^{\prime}\right)\,,
\end{aligned}
\end{equation}
where $x,y,z,a,b$ correspond to the coordinates specified in Eq.~\eqref{xyzab} and
\begin{equation}\label{vav}
 \begin{aligned}
  V_{\text{av}}(R, \theta) & =  \frac{1}{7}[2\left(V(R,\theta,a)+V(R,\theta,b)\right)+ \\
                           & +\left(V(R,\theta,x) +V(R,\theta,y)+V(R,\theta,z)\right)] \,.
 \end{aligned}
\end{equation}
This allowed us to fully include the rotational structure of \ce{H2} in the scattering calculations, thus also incorporating the couplings with the \ce{H2} $(j>0)$ rotational states. However, a dedicate test (see next section for details) showed that the effects of such coupling on the rate coefficients is negligible, as already found for the \ce{HCO+} -- \ce{H2} collisional system \citep{masso2014hco+}. Only the leading term $l_2 = \mu = 0$ needs thus to be retained in the expansion of the interaction potential expressed Eq.~\eqref{fit}.

\begin{figure}
 \includegraphics[width=\columnwidth]{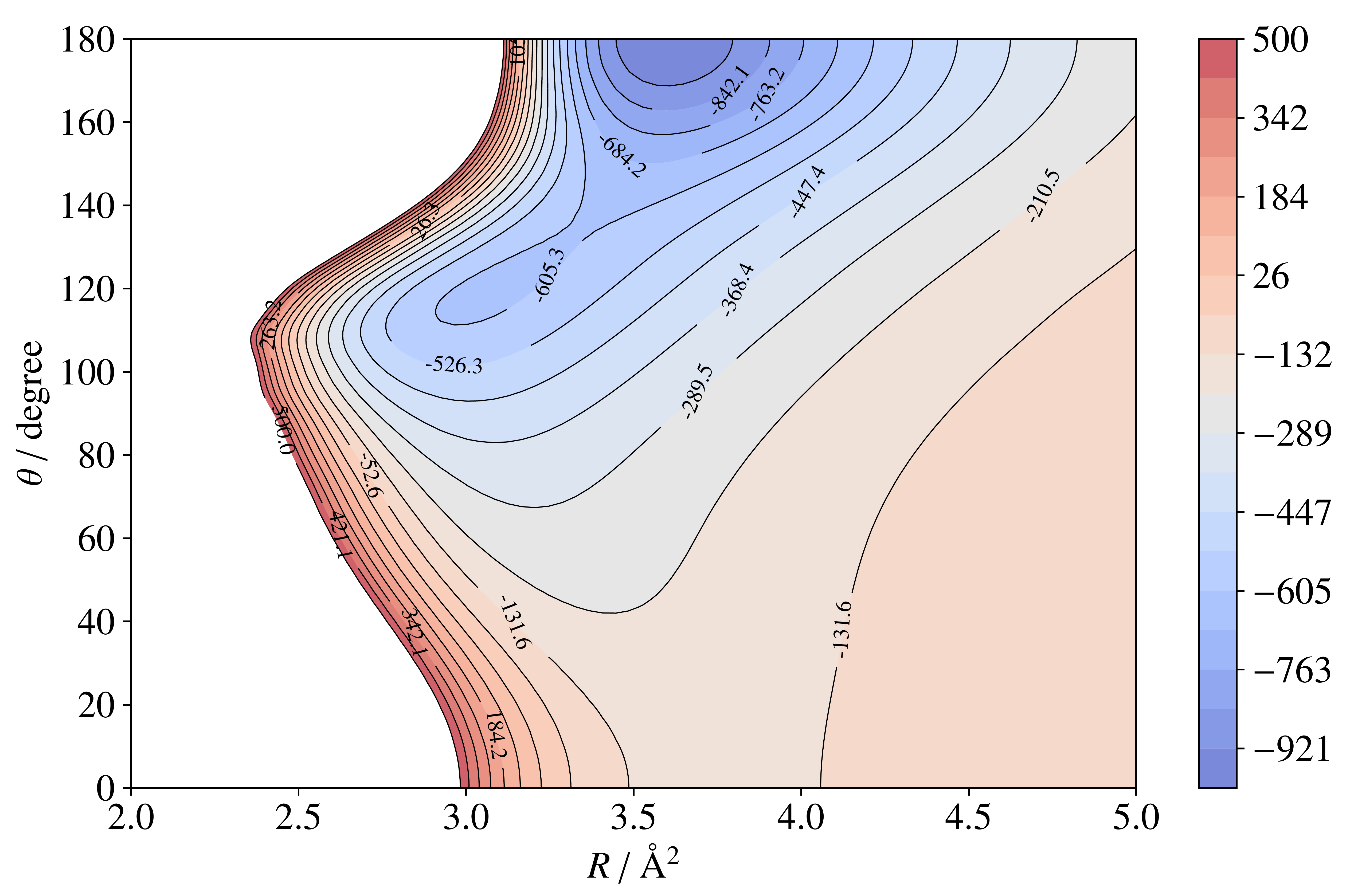}
 \caption{Contour plot of the \ce{HCO+}$-$\,\ce{H2} interaction PES for the averaged potential $V_{\text{av}}$ (Eq. \ref{vav}) as a function of $R$ and $\theta$. Energies are in \wn.}
 \label{fig:pes5av}
\end{figure}

This corresponds to further simplify the expansion of the potential to the interaction of a rigid linear body with a sphere \citep{dumouchel2011rotational,lique2008can}:
\begin{equation}
 V_{\text{av}}(R, \theta)=\sum_{l_1} v_{l_1}(R) P_{l_1}(\cos (\theta)) \,,
\end{equation}
where the quantity $V_{\text{av}}(R,\theta)$ is the potential averaged over the angular motion of \ce{H2}, already defined in Eq.~\eqref{vav}.
This formulation of the potential reduces the description of the PES to a two-dimensional problem, thus significantly simplifying the computational effort of scattering calculations. A contour plot of the potential derived from the fit is shown in Figure \ref{fig:pes5av}.

A spherical average of the \ce{HCO+}--\ce{H2} potential was also previously performed by \citet{yazidi2014revised} but, unlike us, they accounted for only three orientations of \ce{H2}. Comparing the topology of the two PESs, the agreement is good. Both potentials exhibit a strong anisotropy with respect to the $\theta$ angle and present a global minimum at $R = 3.6$\,\AA, and $\theta = 180$ degrees. 
Since we have adopted an higher level of theory, the present PES represents --- to date --- the most accurate description of a spherical potential on the targeted system.

\subsection{Scattering Calculations}\label{scat}
\subsubsection{Inelastic cross sections}\label{cs}

\begin{figure*}
 \includegraphics[scale=0.50]{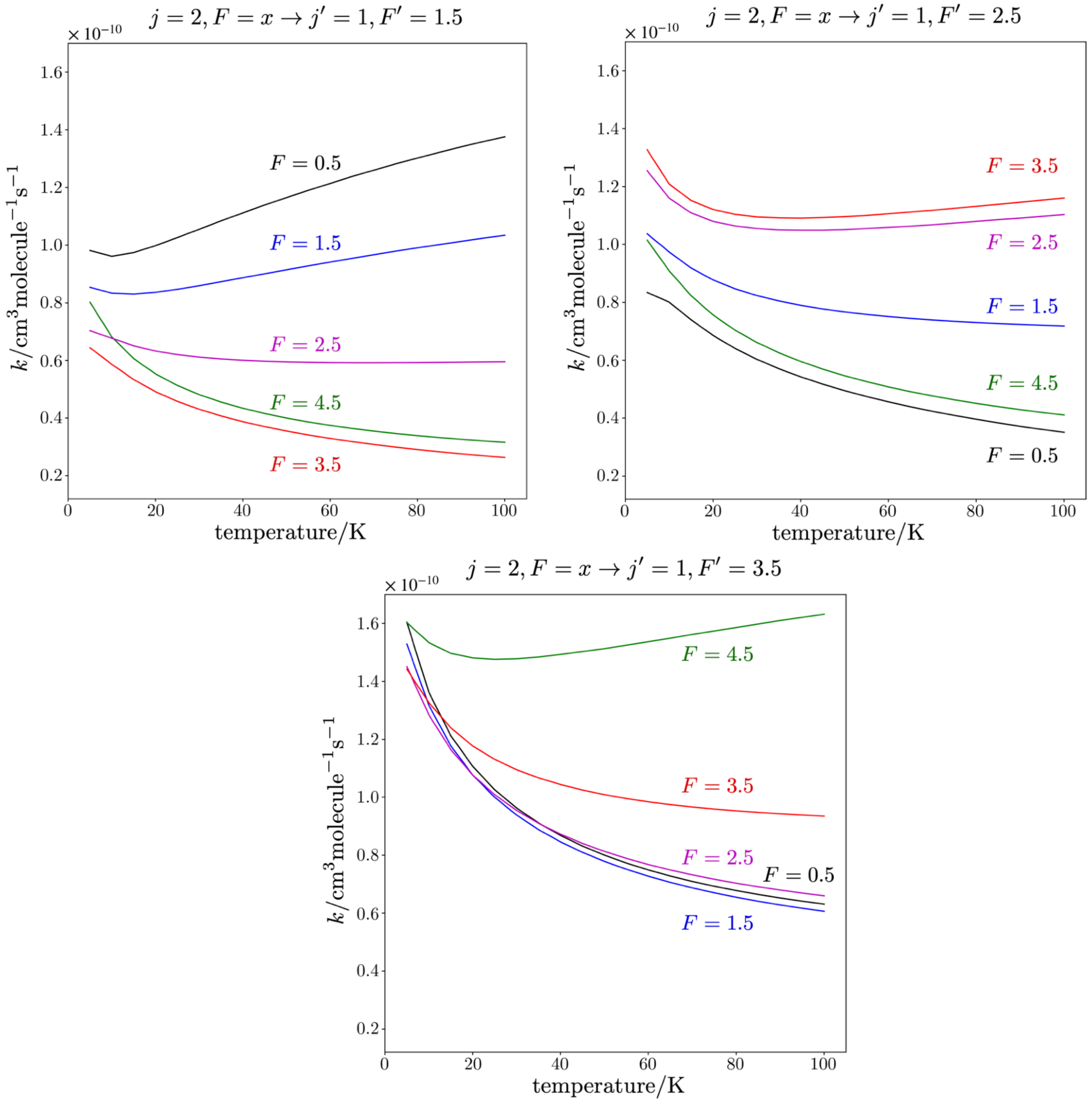}
 \caption{Temperature variation of the hyperfine resolved \ce{HC^{17}O+} -- \ce{H2} rate coefficients for some transitions involving the same final hyperfine state.}
 \label{fig:hfr_inel}
\end{figure*}

\begin{figure}
 \centering
 \includegraphics[scale=0.25]{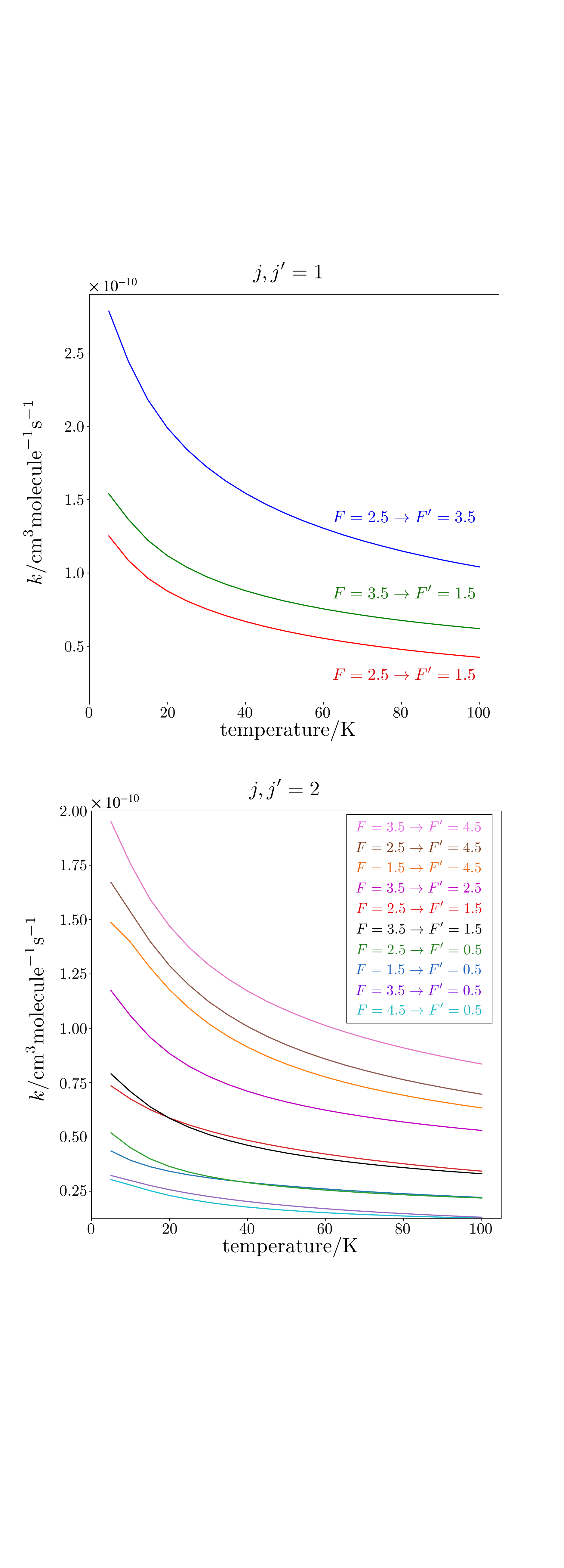}
 \caption{Temperature variation of the hyperfine resolved \ce{HC^{17}O+} -- \ce{H2} quasi-elastic rate coefficients involving the $j=1,2$ rotational levels.}
 \label{fig:hfr_quasiel}
\end{figure}

\indent\indent
With the aim of computing the rate coefficients of the \ce{HC^{17}O+}$-$\,\ce{H2} collisional system, the most reliable approach is to perform full quantum close-coupling (CC) calculations on the interaction potential shifted for the \ce{^{16}O}$\,\rightarrow$\ce{^{17}O} isotope substitution.
This isotopic substitution introduces a new feature due to the \ce{^{17}O} nuclear spin (angular momentum $I=5/2$), which couples with the molecular rotational angular momentum $\mathbf{j}$ through electric quadrupole interaction.
The total angular momentum of the molecule $\mathbf{F}$ is thus defined by the $\mathbf{j} + \mathbf{I}$ vector sum. This leads to the splitting of each rotational level into various $F$ sublevels, with $F$ varying between $\left|I-j\right|$ and $I+j$. 
Since the energy splittings due to the interaction structure are much smaller than those between rotational levels, a common approximation is to ignore, at a first stage, such coupling contribution from the molecular Hamiltonian and to subsequently decouple the spin wave-functions from the rotational ones using the recoupling method described by  \citet{alexander1985collision}.
In detail, the inelastic cross-sections associated with a $jF \rightarrow j'F'$ transition are obtained from the rotational scattering matrix elements $S^{J}(j l ; j' l')$  (nuclear-spin free) derived from the close-coupling (CC) calculations: 
\begin{equation}
T^{J}\left(jl;j'l'\right) = 1 - S^{J}(jl;j'l')\,. 
\end{equation}
$T^{J}\left(jl;j^{\prime}l^{\prime}\right)$ are the $T$-matrix elements, which are linked to the opacities tensor $P^{K}\left(j \rightarrow j^{\prime}\right)$\, via
\begin{equation}
 P^{K}\left(j \rightarrow j^{\prime}\right)=\frac{1}{2 K+1}\,\sum_{l l^{\prime}}\left|\,T^{K}\left(j l ; j^{\prime} l^{\prime}\right)\,\right|^{2}\,,
\end{equation}
where
\begin{equation}
 \begin{aligned}
  T^{K}\left(jl;j^{\prime}l^{\prime}\right)
   &=(-1)^{-j-l^{\prime}}\,(2K+1)\,\sum_{J}(-1)^{J}\,(2J+1) \\
   & \times\left\{
   \begin{array}{ccc}
   l^{\prime} & j^{\prime} & J \\
   j          &    l       & K
   \end{array}\right\}
   \,T^{J}\left(jl;j^{\prime}l^{\prime}\right)\,.
 \end{aligned}
\end{equation}
From the opacity tensor elements $P^{K}\left(j \rightarrow j^{\prime}\right)$, the recoupled inelastic cross sections ($\sigma_{j F \rightarrow j' F'}^{\mathrm{REC}}$) are obtained as:
\begin{equation}
\label{12}
%\begin{aligned}
    \sigma_{j F \rightarrow j^{\prime} F^{\prime}}^{\mathrm{REC}}=
    \frac{\pi}{k_{j F}^{2}}\left(2 F^{\prime}+1\right) \sum_{K}\left\{\begin{array}{ccc}
    j & j^{\prime} & K \\
    F^{\prime} & F & I
    \end{array}\right\}^{2} P^{K}\left(j \rightarrow j^{\prime}\right)\,.
%\end{aligned}  
\end{equation}
This method is approximate, but almost exact and therefore is commonly considered as a "reference" approach (\citet{faure2012impact}). 

\begin{table}
\renewcommand{\tabcolsep}{3pt}
\renewcommand{\arraystretch}{1.1}
    \scriptsize
    \caption{Computed cross sections 
    at $E = 50$\,\wn for \ce{HC^{17}O+}$-$\,\ce{H2}$(j = 0)$ collisions obtained from the 2D spherically averaged potential and the full 4D potential which includes the coupling with $j = 2$ rotational level of \ce{H2}. }
    \footnotesize
    \centering 
    \begin{tabular}{lcddcd}
        \hline\hline \\[-2ex]
        \multirow{2}{*}{$j \rightarrow j^{\prime}$} & & \multicolumn{2}{c}{Cross sections / \AA$^2$} & & \multicolumn{1}{c}{\% Error}\\
        \cline{3-4} \\[-2ex]
        & &  \multicolumn{1}{c}{4D} & \multicolumn{1}{c}{2D}& & \\
        \hline
        1 $\rightarrow$ 0 &  & 58.14   &    53.41      &  &  -8.86    \\                                    
        2 $\rightarrow$ 0 &  & 52.59   &    50.31      &  &  -4.54    \\                                    
        3 $\rightarrow$ 0 &  & 46.12   &    44.00      &  &  -4.84    \\                                    
        4 $\rightarrow$ 0 &  & 27.26   &    30.28      &  &  9.98     \\                                    
        5 $\rightarrow$ 0 &  & 16.81   &    15.74      &  &  -6.80     \\                                                                                              
        2 $\rightarrow$ 1 &  & 59.78   &    60.87      &  &  1.80     \\                                    
        3 $\rightarrow$ 1 &  & 48.43   &    53.18      &  &  8.92     \\
        4 $\rightarrow$ 1 &  & 36.70   &    35.19      &  &  -4.30    \\
        5 $\rightarrow$ 1 &  & 17.09   &    18.77      &  &  8.93     \\                                                                                   
        3 $\rightarrow$ 2 &  & 55.24   &    52.95      &  &  -4.32    \\                                            
        4 $\rightarrow$ 2 &  & 45.75   &    41.75      &  &  -9.56    \\                                            
        5 $\rightarrow$ 2 &  & 20.39   &    21.02      &  &  3.01      \\                                                                                                                               
        4 $\rightarrow$ 3 &  & 56.79   &    51.02      &  &  -11.31   \\                                            
        5 $\rightarrow$ 3 &  & 24.43   &    23.19      &  &  -5.35    \\                                            
        5 $\rightarrow$ 4 &  & 40.04   &    35.63      &  &  -12.39   \\                                              
        \hline
        \multicolumn{4}{l}{Average absolute \% error}  & & 6.99 \\
        \hline\hline
    \end{tabular}
    \label{2d4d}
    \end{table}

As mentioned in the previous section, a preliminary test on the influence of the coupling between $j=0$ and $j>0$ rotational states of \ce{H2} was performed in order to evaluate the reliability of the 2D potential spherically averaged over the \ce{H2} orientations.
We thus compared the values of the $j(\ce{H2}) = 0$ partial cross sections computed for a single value of the total energy ($50$\,\wn) using the global 4D [Eq.~\eqref{fit}], thus accounting for the $j=0,2$ coupling, and the reduced 2D [Eq.~ \eqref{vav}] potentials. 
The results are shown in Table~\ref{2d4d}: the pairs of values exhibit a mean average percentage error of $\sim 7$\% and a maximum discrepancy of 12.4\%. 
This validated the results given by the spherical average potential approximation, which was then applied to the entire energy grid.
In the following, the \ce{H2} projectile is thus considered as a structureless species, behaving as a rotating sphere (\textit{para}-\ce{H2} with $j = 0$). 

All scattering calculations have been carried out by employing the MOLSCAT program \citep{molscat14} at values of the kinetic energy ranging from 2 to 500\,\wn, with narrower steps at low energies (0.2\,\wn) which gradually increase as the energy rises.
The adopted propagator is a hybrid one between the Manolopoulos diabatic modified log-derivative propagator and the Alexander-Manolopoulos-Airy propagator (LDMD/AIRY) \citep{alexander1984hybrid,alexander1987stable,manolopoulos1986improved}. This provides the correct compromise between computational efficiency and accuracy in the description of the potential, since it differently deals with the short- and long-range propagation according to the potential requirements, and thus using narrower propagation steps when the energy gradient is higher.
The integration range was set from 2.5\,\AA\, to a long range limit value purposely chosen to ensure convergence of the inelastic cross sections for the considered energy range. 
The same scheme has been applied to the rotational basis cut-off of \ce{HC^{17}O+}, which has been optimized for each energy range, from a minimum value of $j=25$ at 2\,\wn up to a maximum of $j=31$ at the highest energies (above 300\,\wn). 
The maximum value of the total angular momentum $\mathbf{J} = \mathbf{j} + \mathbf{l}$ has been chosen as the one that allowed for the convergence of the inelastic cross sections within 0.005\,\AA, reaching a maximum value of $J=118$ at 500\,\wn.
The reduced mass of the collisional system is set to 1.8709\,u and the rotational energies have been computed from the experimental ground-state spectroscopic constants \citep{dore2001laboratory}: $B_0$\,(\ce{HC^{17}O+})\,$= 43528.92$\,MHz, \,$D_0$\,(\ce{HC^{17}O+})\,$= 78.96$\,kHz. The energies of the hyperfine levels of \ce{HC^{17}O+} required for the recoupling calculation have been taken from \citet{bizzocchihyp}.

\subsubsection{Hyperfine-resolved rate coefficients}\label{hf}
\begin{table}
\renewcommand{\tabcolsep}{3pt}
\renewcommand{\arraystretch}{1.1}
%\scriptsize
\caption{Hyperfine resolved (de-)excitation rate coefficients of the two lowest rotational levels for $T = 10$, 30 and 50\,K.}
%\vspace{0.5cm}
%\footnotesize
\centering 
%\begin{threeparttable} 
  \begin{tabular}{ll c c c}
    \hline\hline \\[-2ex]
    \multirow{2}{*}{$j,F \rightarrow j^{\prime}, F^{\prime}$}& & \multicolumn{3}{c}{$k_{jF\rightarrow j'F'} / 10^{-10}$cm$^3$ s$^{-1}$} \\[1ex]
    \cline{3-5} \\[-2ex]
    & &  \multicolumn{1}{c}{\hspace{0.5em}10\,K} & \multicolumn{1}{c}{\hspace{0.7em}30\,K} & \multicolumn{1}{c}{50\,K} \\[1ex]
    \hline \\[-2ex]
    1, 1.5 $\rightarrow$ 0, 2.5 &  & 1.85 & 1.54 & 1.45 \\
    0, 2.5 $\rightarrow$ 1, 1.5 &  & 0.81 & 0.89 & 0.89 \\
    1, 2.5 $\rightarrow$ 0, 2.5 &  & 1.85 & 1.54 & 1.45 \\
    0, 2.5 $\rightarrow$ 1, 2.5 &  & 1.22 & 1.34 & 1.34 \\
    1, 2.5 $\rightarrow$ 1, 1.5 &  & 1.09 & 0.75 & 0.60 \\
    1, 1.5 $\rightarrow$ 1, 2.5 &  & 1.63 & 1.13 & 0.91 \\
    1, 2.5 $\rightarrow$ 1, 3.5 &  & 2.45 & 1.73 & 1.41 \\
    1, 3.5 $\rightarrow$ 1, 2.5 &  & 1.83 & 1.29 & 1.06 \\
    1, 3.5 $\rightarrow$ 0, 2.5 &  & 1.85 & 1.54 & 1.45 \\
    0, 2.5 $\rightarrow$ 1, 3.5 &  & 1.63 & 1.78 & 1.78 \\
    1, 3.5 $\rightarrow$ 1, 1.5 &  & 1.37 & 0.98 & 0.81 \\
    1, 1.5 $\rightarrow$ 1, 3.5 &  & 2.73 & 1.95 & 1.62 \\
    \hline \hline
  \end{tabular}
%\end{threeparttable}
\label{rate_coeff}
\end{table}
\indent\indent
Having computed the inelastic cross sections $\sigma_{j F \rightarrow j^{\prime} F^{\prime}}^{\mathrm{REC}}$, the corresponding rate coefficients are straightforwardly derived at a given temperature $T$ by averaging over collision energy ($E_\mathrm{c}$)\.:
\begin{equation}
%\begin{aligned}
 k_{j F \rightarrow j^{\prime} F^{\prime}}(T)=
    \left(\frac{8}{\pi \mu k_{\mathrm{B}}^{3} T^{3}}\right)^{1 / 2} \int_{0}^{\infty} \sigma_{j F \rightarrow j^{\prime} F^{\prime}}^{\mathrm{REC}}\,E_\mathrm{c}
    \,\mathrm{e}^{-E_{\mathrm{c}}/k_{\mathrm{B}T}}\,\mathrm{d} E_{\mathrm{c}}\,.
%\end{aligned}    
\end{equation}
We have obtained hyperfine resolved (de-)excitation rate coefficients for the lowest six rotational levels in the 5$-$100\,K range. 
The complete set of them will be made available through the LAMDA \citep{schoier2010lamda, van2020leiden} and BASECOL \citep{dubernet2013basecol2012} databases. 
A list of the rate coefficients involving the first two rotational levels at 10, 30 and 50\,K is presented in Table~\ref{rate_coeff}, while Figures~\ref{fig:hfr_inel} and~\ref{fig:hfr_quasiel} depict the variation of some de-excitation rate coefficients with temperature. In Figure~\ref{fig:hfr_inel}, the trend of the inelastic rate coefficients involving the same final hyperfine state is provided, whereas Figure~\ref{fig:hfr_quasiel} shows the quasi-elastic rate coefficients into the $j = 1$ and $j = 2$ rotational levels. 
From both, a propensity towards the transitions involving the final hyperfine level ($F'$) with the highest multiplicity clearly stands out.
In addition, inelastic rate coefficients exhibit a propensity toward transitions involving $\Delta F = \Delta j$ when $F'\ge I$ and $-\Delta F = \Delta j$ when $F' < I$.

\begin{figure*}
\centering
	\includegraphics[scale=0.48]{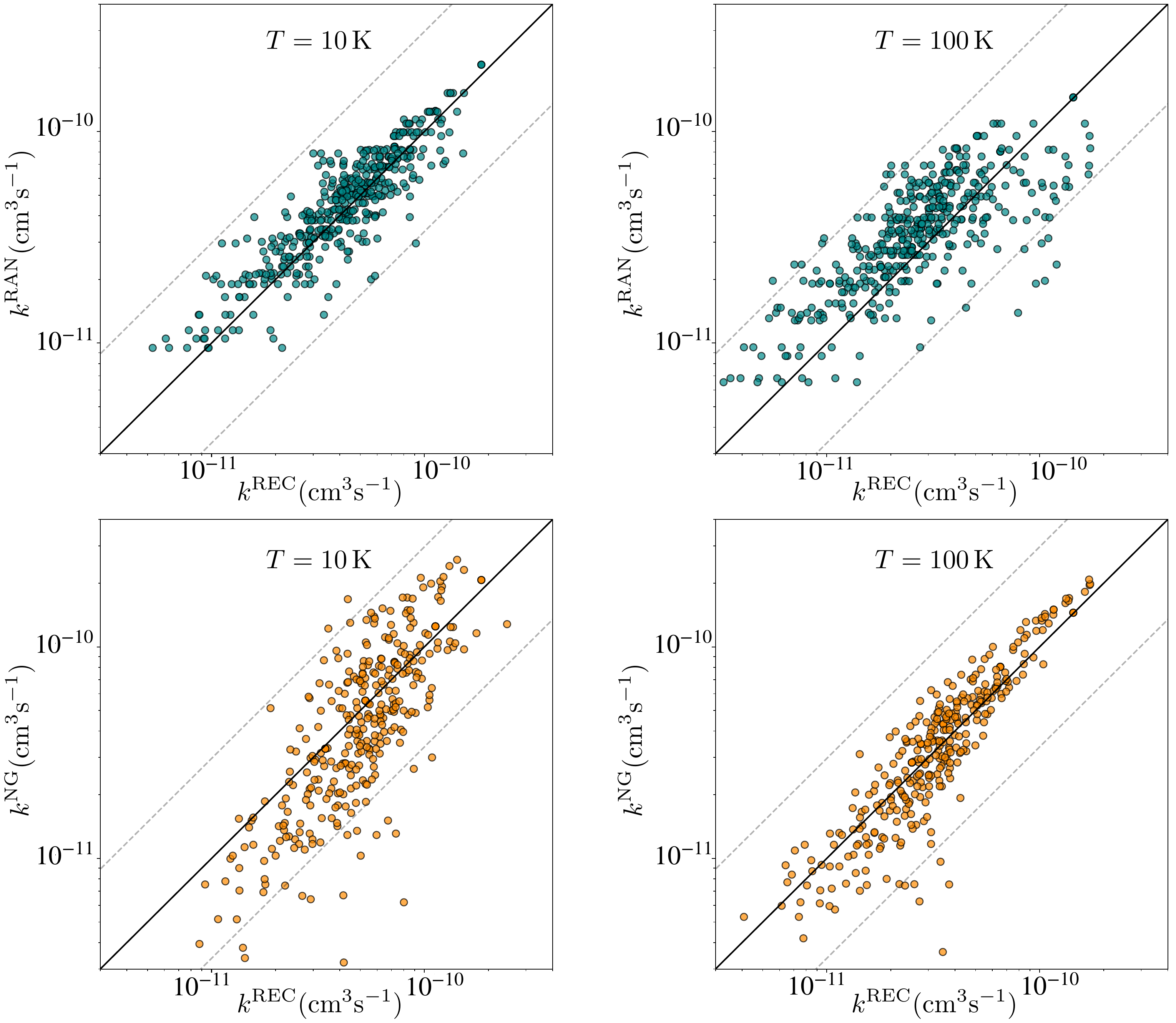}
    \caption{Comparison between \ce{HC^{17}O+} -- \ce{H2} recoupling hyperfine rate coefficients and those obtained using the RAN (top panels) and NG (bottom panels) approximations at 10\,K (left-hand panels) and 100\,K (right-hand panels). In each panel, the dashed lines delimit the region where the rate coefficients differ by less than a factor of 3.}
    \label{fig:cc_ios_RAN}
\end{figure*}

\section{Isotopic effect}\label{ie}
\indent\indent
For astrophysical purposes, it is often assumed that collisional rate coefficients of a main isotopologue can be used to estimate rate coefficients for other isotopic variants. 
To test the reliability of this approximation for the \ce{^{16}O}$\,\rightarrow$\ce{^{17}O} isotope substitution, we compared the values of the cross section computed at $E = 50$\,\wn for the \ce{HCO+} and \ce{HC^{17}O+} target species. 
These two scattering system differ by the position of the centre of mass, the rotational constants of the target and the reduced masses. 
The results, reported in Table~\ref{1716}, exhibited significant discrepancies, with an average percentage difference of more than 20\%. Noteworthy, the variation between the cross sections of \ce{HCO+} and \ce{HC^{17}O+} seems not to follow any kind of predictable pattern. Therefore we expect that, the use of scaled rate coefficients derived from the parent species, would lead to unreliable results in radiative transfer modelling of astrophysical \ce{HC^{17}O+} emission. 

\begin{table}
\renewcommand{\tabcolsep}{3pt}
\renewcommand{\arraystretch}{1.1}
    \scriptsize
    \caption{Computed cross-sections 
    at $E = 50$\,\wn for the \ce{HC^{17}O+}/\ce{HCO+} and \ce{H2}$(j = 0)$ collisions.}
    %\vspace{0.5cm}
    \footnotesize
    \centering 
    \begin{threeparttable} 
    \begin{tabular}{lcddcd}
        \hline\hline \\[-2ex]
        \multirow{2}{*}{$j \rightarrow j^{\prime}$} & & \multicolumn{2}{c}{Cross sections / \AA$^2$} & & \multicolumn{1}{c}{\% Error} \\ [0.5ex]      
        \cline{3-4}  
       
       \addlinespace[1ex] & &  \multicolumn{1}{c}{\ce{HC^{17}O+} } & \multicolumn{1}{c}{\ce{HCO+} }& & \\
        \midrule                                
        1 $\rightarrow$ 0 &  &   53.41        &       64.96          &  &         21.62                \\ 
        2 $\rightarrow$ 0 &  &   50.31        &       55.34          &  &         10.00                \\
        3 $\rightarrow$ 0 &  &   44.00        &       47.18          &  &         7.24                 \\     
        4 $\rightarrow$ 0 &  &   30.28        &       41.16          &  &         35.90                \\                
        5 $\rightarrow$ 0 &  &   15.74        &       20.49          &  &         30.20                \\        
        2 $\rightarrow$ 1 &  &   60.87        &       60.25          &  &         -1.03                \\           
        3 $\rightarrow$ 1 &  &   53.18        &       49.24          &  &         -7.40                \\
        4 $\rightarrow$ 1 &  &   35.19        &       40.20          &  &         14.22                \\
        5 $\rightarrow$ 1 &  &   18.77        &       18.88          &  &         0.59                 \\
        3 $\rightarrow$ 2 &  &   52.95        &       64.27          &  &         21.39                \\
        4 $\rightarrow$ 2 &  &   41.75        &       51.14          &  &         22.47                \\  
        5 $\rightarrow$ 2 &  &   21.02        &       23.86          &  &         13.47                \\
        4 $\rightarrow$ 3 &  &   51.02        &       61.93          &  &         21.38                \\                  
        5 $\rightarrow$ 3 &  &   23.19        &       37.58          &  &         62.02                \\                              
        5 $\rightarrow$ 4 &  &   35.63        &       47.33          &  &         32.85 \\                                                         
        \hline
        \multicolumn{4}{l}{Average absolute \% error}  & & 20.12 \\
        \hline\hline
    \end{tabular}
    \end{threeparttable}
    \label{1716}
    \end{table}

\section{Comparison with approximated rate coefficients}\label{ap}
\indent\indent
The recoupling approach used in the present work to compute hyperfine-resolved rate coefficients is almost exact but requires to store the $S^{J}(j l ; j' l')$ elements and to compute the opacity tensor between the rotational levels of \ce{HC^{17}O+}, which implies a significant computational effort.
To avoid such a demanding step, various approximate methodologies are frequently employed.
A simple approach that is widely used for astrophysical applications (see \citealt{guilloteau1981thermal,keto2010modeling})
is the statistical method called $M_j$-randomizing limit or proportional approach \citep[RAN,][]{alexander1985collision}. 
This method assumes a statistical reorientation of the quantum number $F'$ after the collision, thus neglecting any dependence on the initial conditions of the system.
This allows to express the $jF\rightarrow j'F'$ hyperfine-resolved rate coefficients from the corresponding pure rotational one $j\rightarrow j'$ through
\begin{equation}
  k_{jF\rightarrow j'F'}(T) = \frac{(2F'+1)}{(2j'+1)(2I+1)} k_{j\rightarrow j}(T) \,.
\end{equation}
A way to account for the collision propensities given by the Wigner coefficients of Eq.~\eqref{12}, is provided by the \citet{neufeld1994excitation} approximation (NG). This approach is based on the IOS method which ignores the rotational energy spacing with respect to the collision energy, and derives the rate coefficients between hyperfine structure levels from the rotational excitation rate involving the fundamental $j=0$ state (for more in-depth description, see \citealt{faure2012impact} and references therein).

To assess the impact of these approximations on a typical non-LTE radiative transfer modelling, we have compared the hyperfine rate coefficients obtained with the full quantum CC approach plus recoupling with those derived from the corresponding rotational collision data employing the NG and RAN approximations. The results at 10\,K and 100\,K are shown in Figure \ref{fig:cc_ios_RAN}. 
At low temperature, the collisions are slow and typically go through the formation of a long lifetime \ce{HC^{17}O+}--\ce{H2} complex in the potential well. When the complex dissociates, the level population evolves to a randomized distribution according to the statistical weights of the final states ($F^{\prime}$). For this reason, the RAN approximation performs better at low temperatures, at which statistical effects have a major impact on the rate coefficients. Conversely, the NG method is less accurate in describing the rate coefficients at low temperatures, where the assumption of neglecting rotational energy spacing compared to the collision energy is no longer valid. On the other hand, at higher temperatures, the influence of the propensity rules given by the Wigner coefficients becomes increasingly prominent, thus making the NG approach a better approximation.

\section{Conclusions}
\indent\indent
In this work, we presented the first computed rate coefficients for the hyperfine (de-)excitation of \ce{HC^{17}O+} by collisions with \ce{H2} ($j=0$), the most abundant collisional partner in cold molecular clouds.
First, we characterized the involved PES by exploiting the accuracy of explicitly correlated coupled-cluster calculations, employing the CCSD(T)-F12a/aug-cc-pVQZ level of theory. The interaction energy was averaged over five \ce{H2} orientations and then fitted by means of the procedure described by \citet{werner1989quantum}. Before performing scattering calculations, the effects due to the coupling between the different rotational states of \ce{H2} ($j = 0,2$) on the inelastic cross sections have been assessed. This permitted us to neglect, in a good approximation, the influence of $j(\ce{H2}) > 0$, thus allowing to perform spherical average of the potential with respect to the orientations of \ce{H2}. Finally, state-to-state rate coefficients between the six lowest rotational levels have been computed using recoupling techniques for temperatures ranging from~5 to 100\,K.

The importance of these data is highlighted by the significant difference between the values of the collision cross sections computed for \ce{HCO+} and \ce{HC^{17}O+}. 
As a matter of facts, retrieving the collisional rate coefficients of \ce{HC^{17}O+} and \ce{H2} by simply scaling the ones of the \ce{HCO+}--\ce{H2} system does not represent a reliable approach. A similar behaviour is also expected for the \ce{HC^{18}O+} isotopologue, which features an even more pronounced shift of the centre of mass with respect to the parent species. 
Furthermore, the comparison with commonly adopted approaches, such as RAN and NG approximations, indicates that the recoupling approach represents the most reliable methodology to compute hyperfine resolved inelastic rate coefficients for astrophysically interesting systems.

\section*{Acknowledgements}
This study was supported by MUR (PRIN Grant Number 202082CE3T) and University of Bologna (RFO funds). The SMART@SNS Laboratory (http://smart.sns.it) is acknowledged for providing high-performance computing facilities.
We also acknowledge financial support from the European Research Council (Consolidator Grant COLLEXISM, Grant Agreement No. 811363).
F.~Lique acknowledges financial support from the Institut Universitaire de France and the Programme National ``Physique et Chimie du Milieu Interstellaire'' (PCMI) of CNRS/INSU with INC/INP cofunded by CEA and CNES.

%%%%%%%%%%%%%%%%%%%%%%%%%%%%%%%%%%%%%%%%%%%%%%%%%%
\section*{Data Availability}

The data underlying this article will be made available through the LAMDA \citep{schoier2010lamda, van2020leiden} and BASECOL \citep{dubernet2013basecol2012} databases.



%\bibliographystyle{mnras}
%\bibliography{hc17o+_paper} % if your bibtex file is called references.bib

%\begin{thebibliography}{99}
%\end{thebibliography}

%%%%%%%%%%%%%%%%%%%%%%%%%%%%%%%%%%%%%%%%%%%%%%%%%%

%%%%%%%%%%%%%%%%% APPENDICES %%%%%%%%%%%%%%%%%%%%%

%\appendix
%
%\section{Some extra %material}
%
%If you want to %present additional %material which would %interrupt the flow of %the main paper,
%it can be placed in %an Appendix which %appears after the %list of references.

%%%%%%%%%%%%%%%%%%%%%%%%%%%%%%%%%%%%%%%%%%%%%%%%%%

% Don't change these lines
\bsp	% typesetting comment
\label{lastpage}
\end{document}